\documentclass[prl,twocolumn,superscriptaddress,showpacs,floatfix]{revtex4-1}
\usepackage{amssymb}
\usepackage{graphicx}
\usepackage{subfigure}	 	
\usepackage[english]{babel}
\usepackage{float}
\usepackage{color}
\usepackage{bm}
\usepackage{dcolumn}
\usepackage{amsmath}

\usepackage{epsfig}

\begin{document}

\title{Dissipative vortex solitons in 2D-lattices}

\author{C. Mej\'ia-Cort\'es}
\affiliation{Instituto de \'Optica, C.S.I.C., Serrano 121, 28006 Madrid, Spain}

\author{J.M. Soto-Crespo}
\affiliation{Instituto de \'Optica, C.S.I.C., Serrano 121, 28006 Madrid, Spain}

\author{Mario I. Molina}
\affiliation{Departamento de F\'isica, Facultad de Ciencias, Universidad de Chile, Casilla 653, Santiago, Chile}
\affiliation{Center for Optics and Photonics, Universidad de Concepci\'on, Casilla 4016, Concepci\'on, Chile}

\author{Rodrigo A. Vicencio}
\affiliation{Departamento de F\'isica, Facultad de Ciencias, Universidad de Chile, Casilla 653, Santiago, Chile}
\affiliation{Center for Optics and Photonics, Universidad de Concepci\'on, Casilla 4016, Concepci\'on, Chile}

\date{\today}

\begin{abstract}
We report the existence of stable symmetric vortex-type solutions for
two-dimensional nonlinear discrete dissipative systems governed by a
cubic-quintic complex Ginzburg-Landau equation. We construct a whole family of
vortex solitons with a topological charge $S=1$. Surprisingly, the dynamical
evolution of unstable solutions of this family does not alter significantly
their profile, instead their phase distribution completely changes.  They
transform into \textit{two-charges swirl-vortex solitons}. We dynamically excite
this novel structure showing its experimental feasibility.
\end{abstract}


\maketitle

The study of discrete nonlinear systems has been an active area of research
during the last twenty years due to its broad impact in diverse branches of
science and to its potential for technological
applications~\cite{PT,rep1,rep2,chrinat}. Until now, nonlinear optics has been
the main scenario chosen to test this phenomenon, essentially because of both,
its comparative experimental simplicity and its direct connection with
theoretical models. Nonlinear self-localized structures, usually termed as
discrete solitons, have been predicted and observed for one- and two-dimensional
arrays~\cite{heis,fleis}. A discrete vortex soliton is defined as a nonlinear
self-localized structure whose phase changes $2\pi S$ radians azimuthally. $S$
is an integer number known as the vorticity or topological charge of the
solution. The existence of discrete vortex solitons in conservative systems have
been reported on several works~\cite{vortex1}. For the continuous case,
dissipative vortex soliton families have been found to be stable for a wide
interval of $S$-values~\cite{opex09}. Very recently, symmetric stable vortices
have also been predicted in continuous dissipative systems with a periodic
linear modulation~\cite{contiperio}. 

Nowadays, dissipative models offer a more
complete and realistic description of different physical systems. In
conservative models, gain and loss are completely neglected and the dynamical
equilibrium is reached by means of a balance between nonlinear and dispersive
effects. For dissipative systems, it must also exist an additional balance
between gain and losses, turning the equilibrium into a more complex
process~\cite{soto}. The Ginzburg-Landau equation is - somehow - a universal
model where dissipative solitons are their most interesting solutions. This
model appears in many branches of science like, for example, nonlinear optics,
Bose-Einstein condensates, chemical reactions, superconductivity and many
others~\cite{akhm0508}.

In this work we deal with discrete vortex solitons in dissipative 2D lattices
governed by a discrete version of the Ginzburg-Landau equation. We have found
different families of these localized solutions connected successively by means
of saddle-node bifurcations. We studied their stability and found two types of
stable vortex families coexisting for the same set of parameters. We have
dynamically unveiled the second type of stable solution by following the
decaying of an initially unstable vortex. This observation is very different to
the results shown in Ref.~\cite{contiperio}, where an unstable vortex just
vanishes on propagation, through completely radiative decay. Moreover, our final
vortex solution possesses a nontrivial phase structure where two different
charges coexist.  

Beam propagation in 2D dissipative waveguide lattices can be modeled by the following
equation:
\begin{eqnarray}
\nonumber
&i\dot\psi_{m,n}+\hat C\psi_{m,n}+|\psi_{m,n}|^{2}\psi_{m,n}+\nu|\psi_{m,n}|^{4}\psi_{m,n}=\\
&i\delta\psi_{m,n}+i\varepsilon|\psi_{m,n}|^{2}\psi_{m,n}+i\mu|\psi_{m,n}|^{4}\psi_{m,n}\ .
\label{dgl2}
\end{eqnarray}
Eq.(\ref{dgl2}) represents a physical model for open systems that exchange
energy with external sources and it is called $(2+1)$ discrete complex
cubic-quintic Ginzburg-Landau equation. $\psi_{m,n}$ is the complex field
amplitude at the $(m,n)$ lattice site and $\dot\psi_{m,n}$ corresponds to its
first derivative respect to the propagation coordinate $z$. The set
$\{m=1,...,M\}\times\{n=1,...,N\}$ defines the array, being $N$ and $M$ the
number of sites in the horizontal and vertical directions (in all our
computations $N=M=17$). The fields propagating in each waveguide interact only
with nearest-neighbors through their evanescent tails. This interaction is
described by the discrete diffraction operator $\hat
C\psi_{m,n}=C(\psi_{m+1,n}+\psi_{m-1,n}+\psi_{m,n+1}+\psi_{m,n-1})$, where $C$
is a complex number. Its real part indicates the strength  of the coupling
between different sites and its imaginary part denotes the gain or loss
originated by this coupling. The nonlinear higher order Kerr term is represented
by $\nu$ while $\varepsilon>0$ and $\mu < 0$ are the coefficients for cubic gain
and quintic losses, respectively. Linear losses are determined by negative
$\delta$. 

Unlike the conservative discrete nonlinear Schr$\ddot{\text{o}}$dinger (DNLS)
equation, the power defined as
\begin{equation}
Q(z)=\sum_{m,n=1}^{M,N}\ |\psi_{m,n}(z)|^2,
\end{equation}
is not a conserved quantity in the present model. However, for a self-localized
solution, the power and its evolution will be the main magnitude that we will
monitor in order to identify different families of stationary solutions.

We look for stationary solutions of Eq.(\ref{dgl2}) of the form
$\psi_{m,n}(z)=\phi_{m,n}\exp [i \lambda z]$ where $\phi_{m,n}$ are complex
numbers and $\lambda$ is real; also we are interested that the phase of
solutions change azimuthally an integer number ($S$) of $2\pi$. In such a case
the self-localized solution is called a discrete vortex soliton~\cite{kevre2005}
with vorticity $S$. By inserting the previous {\em ansatz} into model
(\ref{dgl2}) we obtain the following set of algebraic coupled equations:
\begin{eqnarray}
\nonumber
&-\lambda\phi_{m,n}+\hat C\phi_{m,n}+|\phi_{m,n}|^{2}\phi_{m,n}+\nu|\phi_{m,n}|^{4}\phi_{m,n}=\\
&i\delta\phi_{m,n}+i\varepsilon|\phi_{m,n}|^{2}\phi_{m,n}+i\mu|\phi_{m,n}|^{4}\phi_{m,n}\ .
\label{adgl2}
\end{eqnarray}
We look for vortex-type solutions by solving equations (\ref{adgl2}) with a
multi-dimensional Newton-Raphson iterative algorithm. The method requires an
initial guess that we construct as explained below.

\begin{figure}
\centering
\epsfig{file=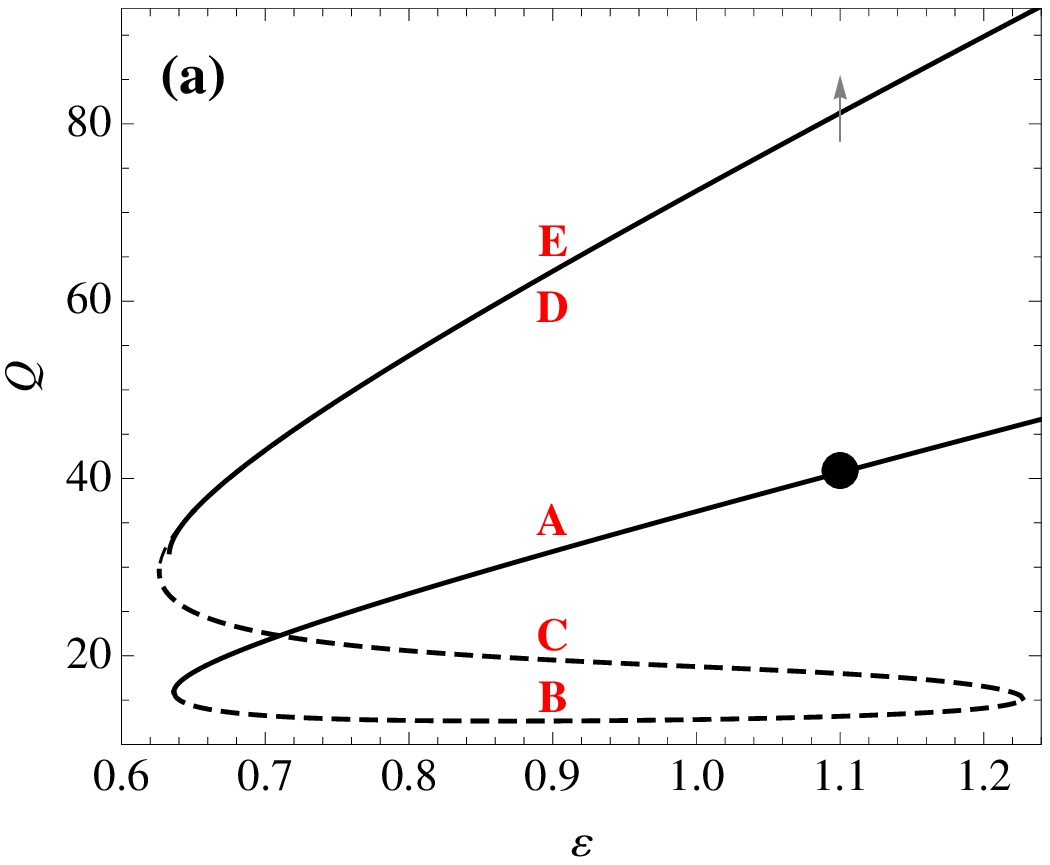,width=0.7\linewidth,clip=}
\begin{tabular}{cc}
\epsfig{file=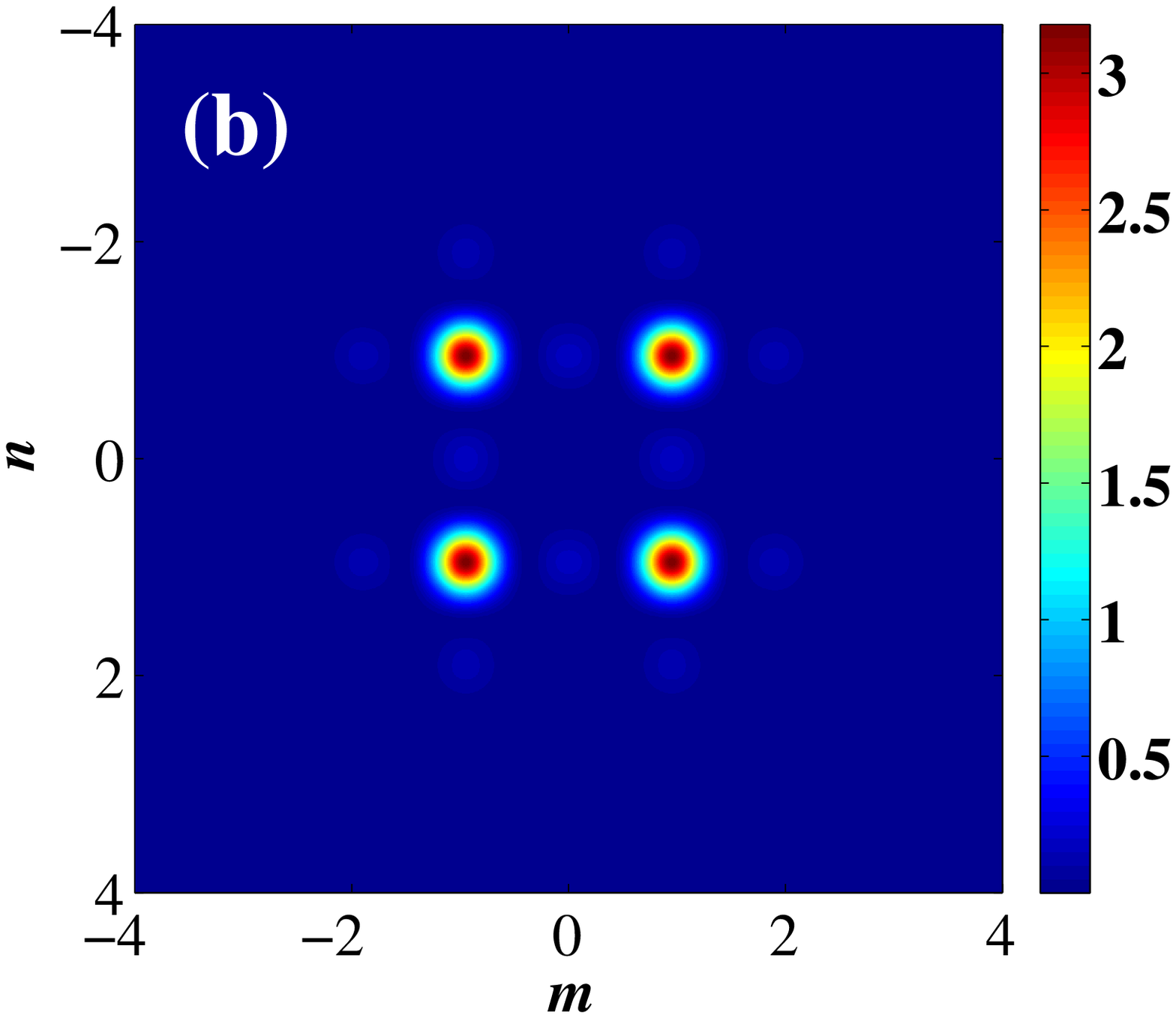,width=0.515\linewidth,clip=}&
\epsfig{file=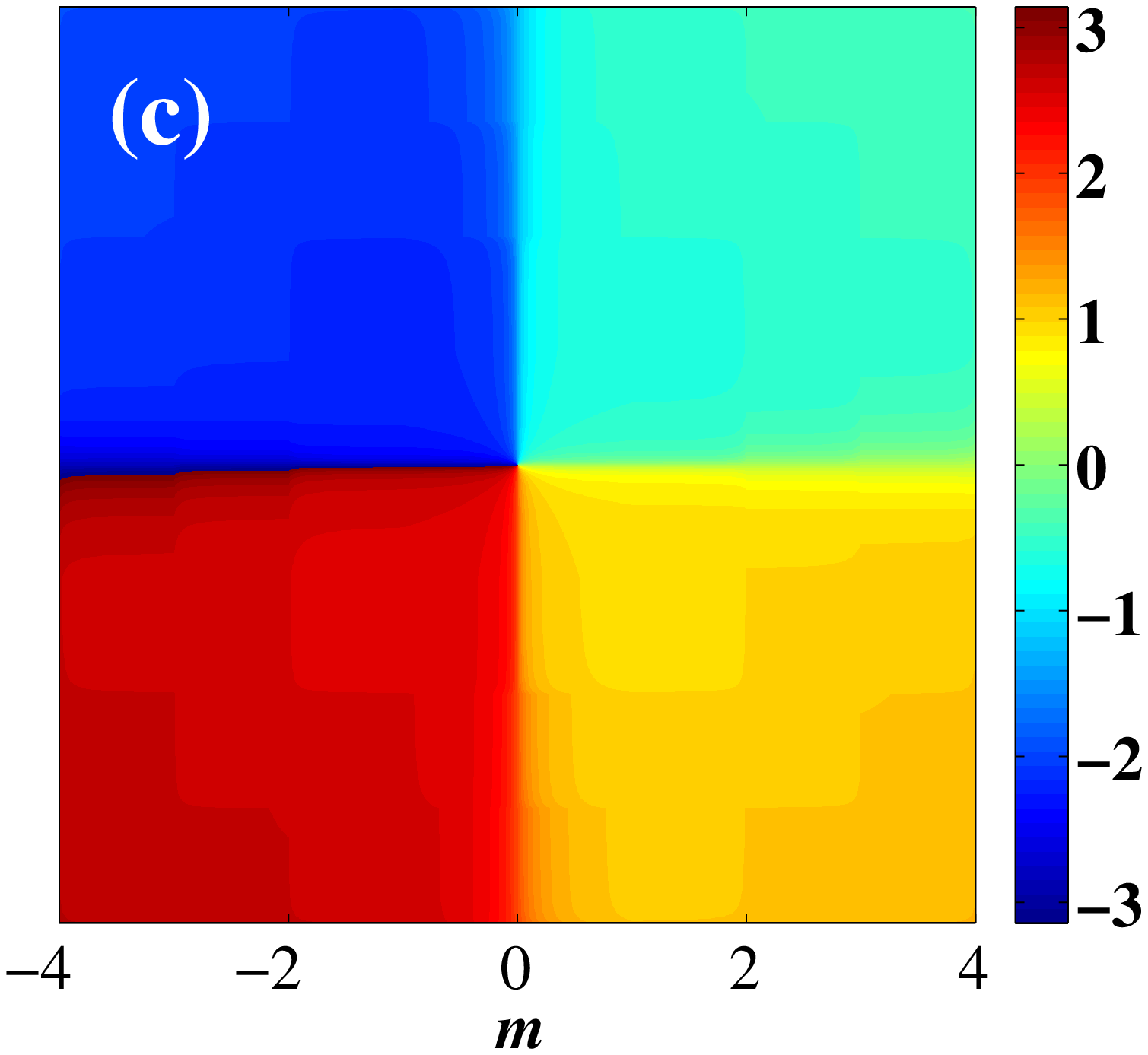,width=0.46\linewidth,clip=}
\end{tabular}
\caption{(Color online) (a) $Q$ versus $\varepsilon$ diagram for discrete vortex
solitons. Continuous and dashed lines correspond to stable and unstable
solutions, respectively. (b) and (c) Color map plots for the amplitude and phase
profiles, respectively, of the solution for $\varepsilon=1.1$ indicated by a black
dot in (a). $C=0.8$, $\delta=-0.9$, $\mu=-0.1$, $\nu=0.1$.}
\label{fig1}
\end{figure}

In the high-confinement limit, single peak solutions (fundamental bright
solitons) were predicted to exist in dissipative nonlinear media \cite{efre07}.
In that limit, we obtain the following approximation:
$\phi_{0}^{2}\approx-(\varepsilon+\sqrt{\varepsilon^{2}-4\mu\delta})/(2\mu)$,
$\lambda\approx\phi_{0}^{2}+\nu\phi_{0}^{4}$, and
$\alpha\approx|C\phi_{0}/(\lambda+i\delta)|$. Here, $\phi_{0}$ corresponds to
the central amplitude, $\lambda$ the nonlinear propagation constant, and
$\alpha$ the first adjacent amplitudes. We can see from the above that the
amplitude of each peak is a function of $\varepsilon$, $\mu$ and $\delta$; if we
set the last two parameters, the amplitude takes a $bi$-quadratic form with
$\varepsilon$ as the bifurcation parameter. Now, we place a single peak
approximation at each corner of a square sub-lattice $L$ as a superposition of
four fundamental bright solitons~\cite{kivshar2005}:
\begin{align}
L &=
\begin{bmatrix}
0 \quad&\quad  \alpha \quad &\quad 0 \quad &\quad\alpha \quad &\quad 0 \\
\alpha\quad&\quad\phi_{0}\quad&\quad\tilde\alpha \quad&\quad\phi_{0} \quad&\quad\alpha\\
0 \quad&\quad  \tilde\alpha \quad &\quad 0 \quad &\quad\tilde\alpha \quad &\quad 0 \\
\alpha\quad&\quad\phi_{0}\quad&\quad\tilde\alpha \quad&\quad\phi_{0} \quad&\quad\alpha\\
0 \quad&\quad  \alpha \quad &\quad 0 \quad &\quad\alpha \quad &\quad 0 \\
\end{bmatrix},\label{init} \\ \nonumber
\label{inita}
\end{align}
where $\tilde\alpha=2\alpha$. Now, we define a phase operator $\Theta$ as
$\Theta_{m,n}=\exp\{i{[\arctan({-n/m})]}\}$, and write our initial $S=1$ {\em
ansatz} as
\begin{equation*}
\phi_{m,n}=L_{m,n}\cdot\Theta_{m,n}\ .
\label{semilla}
\end{equation*}
With this initial guess, we construct a family of 4-peaks symmetric vortex
solitons with vorticity $S=1$, whose stability is monitored through a standard
linear stability analysis~\cite{stabi}. Figure \ref{fig1}(a) shows a $Q$ versus
$\varepsilon$ diagram for these solutions including their stability. This figure
shows the coexistence, for the same set of parameters, of two different branches
of stable solutions and, also, three different families of unstable solitons.
Different families are successively connected by saddle-node  bifurcation
points. An example for a solution of branch A is shown in Figs.\ref{fig1}(b) and
(c) [black dot in Fig.\ref{fig1}(a)].  

This solution is very similar to our initial {\em ansatz} sketched in
Eq.(\ref{init}) with a full topological charge $S=1$. This agreement validates
the seed we constructed as a first approach to find stationary vortex-type
solutions. As the nonlinear amplification is diminished, the stable branch $A$
reaches a first saddle-node point for $\varepsilon\approx0.637$. At this point,
this family turns around and a new family emerges: the unstable branch labeled
$B$. After that, two more saddle-node points appear connecting the new branches
$B$ with $C$ and, then, $C$ with $D$. The unstable branch $D$ is mostly hidden
because it is located at the same region that the stable branch labeled $E$.
Branches $A-D$ preserve the vorticity $S=1$ while the amplitude profiles change
adiabatically. It is worth mentioning that branches $A$, $D$ and $E$ also exist
for higher $\varepsilon$ values, with the power increasing monotonically, as the
high-confinement limit predicts.

\begin{figure}
\centering
\begin{tabular}{cc}
\epsfig{file= 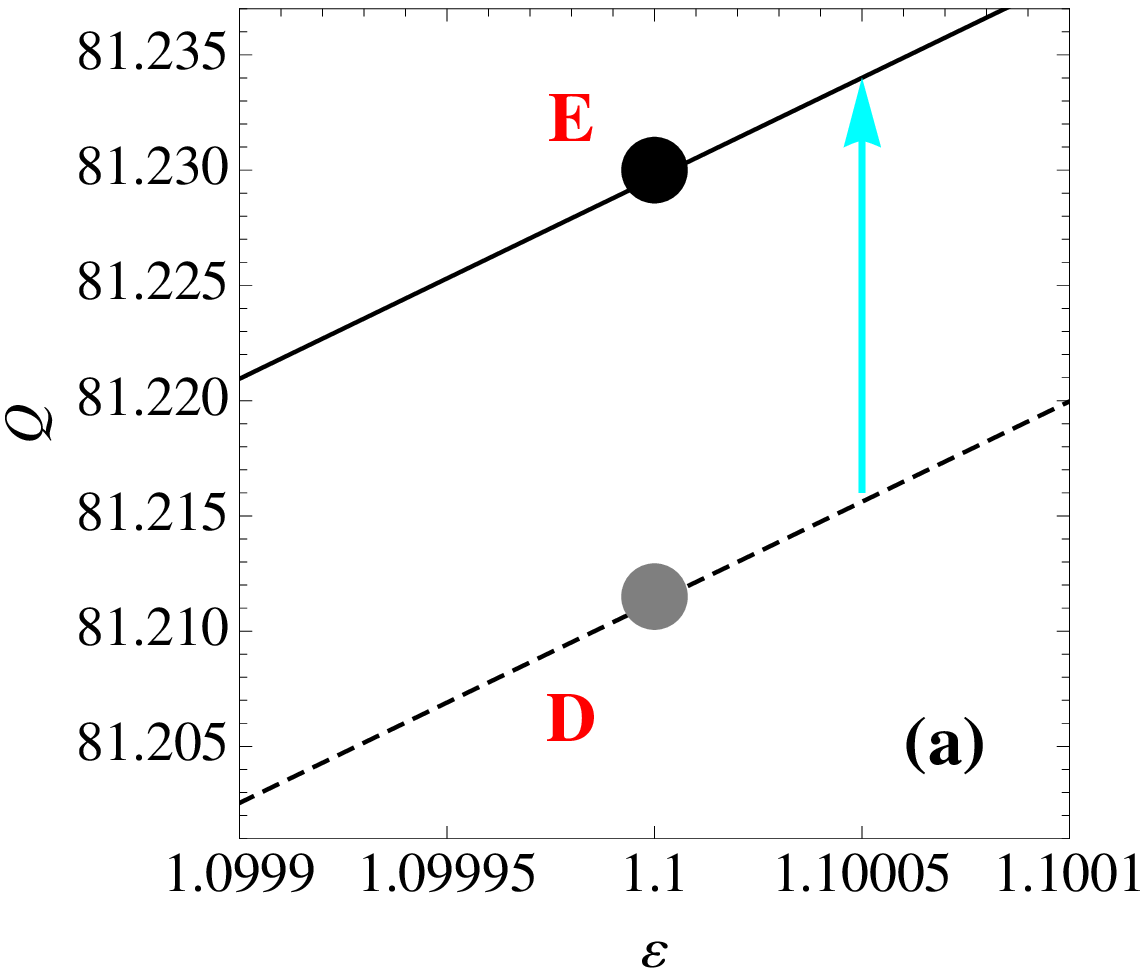,width=0.52\linewidth,clip=}&
\epsfig{file= 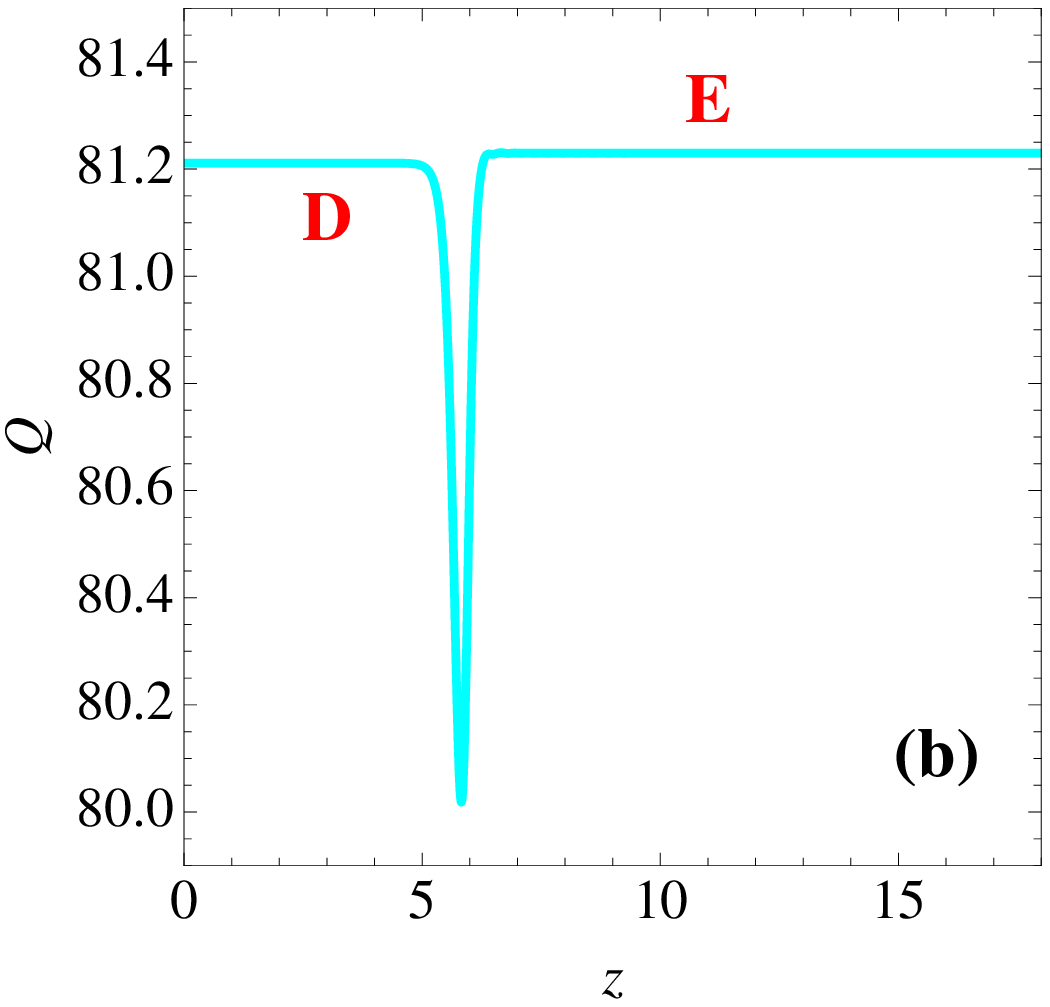,width=0.46\linewidth,clip=}
\end{tabular}
\caption{(Color online) (a) Zoom of branches $D$ and $E$ for a narrow region
around the gray arrow in the Fig.\ref{fig1}(a). (b) Numerical simulation
of model (\ref{dgl2}) showing the power transition sketched by an cyan arrow
in (a) for $\varepsilon=1.1$.}
\label{fig2} 
\end{figure}

As said before, in Fig.\ref{fig1}(a), curves $D$ and $E$  are indistinguishable.
In order to see their differences more clearly, we plot a zoom in
Fig.\ref{fig2}(a) of region $Q\sim 81.2$ for a narrow region around the gray
arrow in the Fig.\ref{fig1}(a). The first solution on branch $E$ [black dot in
Fig.\ref{fig2}(a)] was obtained dynamically; i.e., we numerically integrated
Eq.(\ref{dgl2}) by using an unstable solution [gray dot in Fig.\ref{fig2}(a)] as
initial condition. Contrary to previous observations, for the evolution of
unstable vortex solitons~\cite{contiperio}, we noticed that the power $Q$ makes
one oscillation and then stabilizes very rapidly around a new equilibrium value
[see Fig.\ref{fig2}(b)]. This new value was indeed very close to the initial
one, but now it corresponds to a new stationary solution that propagates stably
by keeping the same amplitude profile but a different phase structure. We took
this new solution as an initial guess in our Newton-Raphson scheme and we
constructed the whole stable branch $E$ shown in Fig.\ref{fig1}(a).

\begin{figure}
\centering
\begin{tabular}{ccc}
\epsfig{file=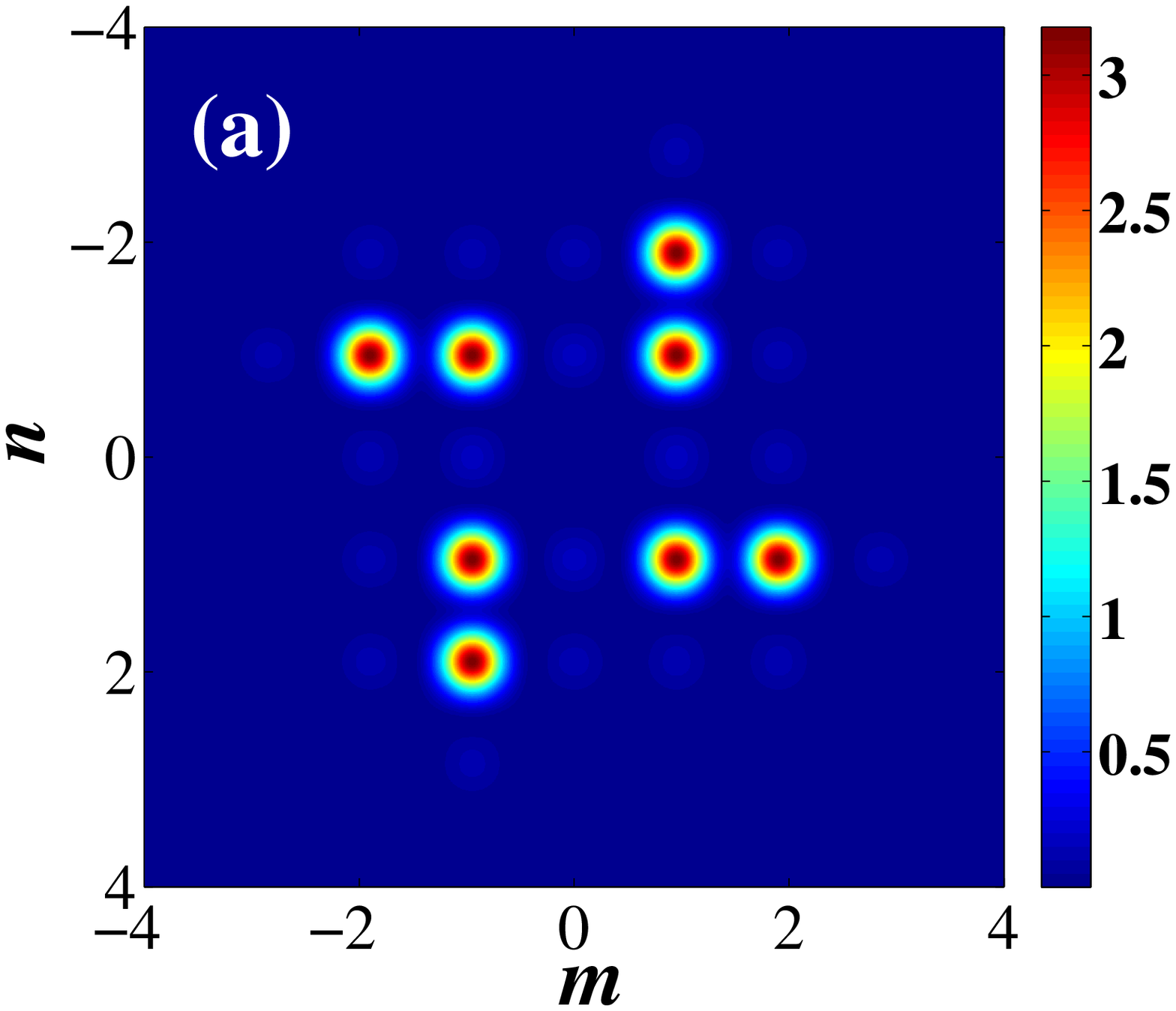,width=0.37\linewidth,clip=}&
\epsfig{file=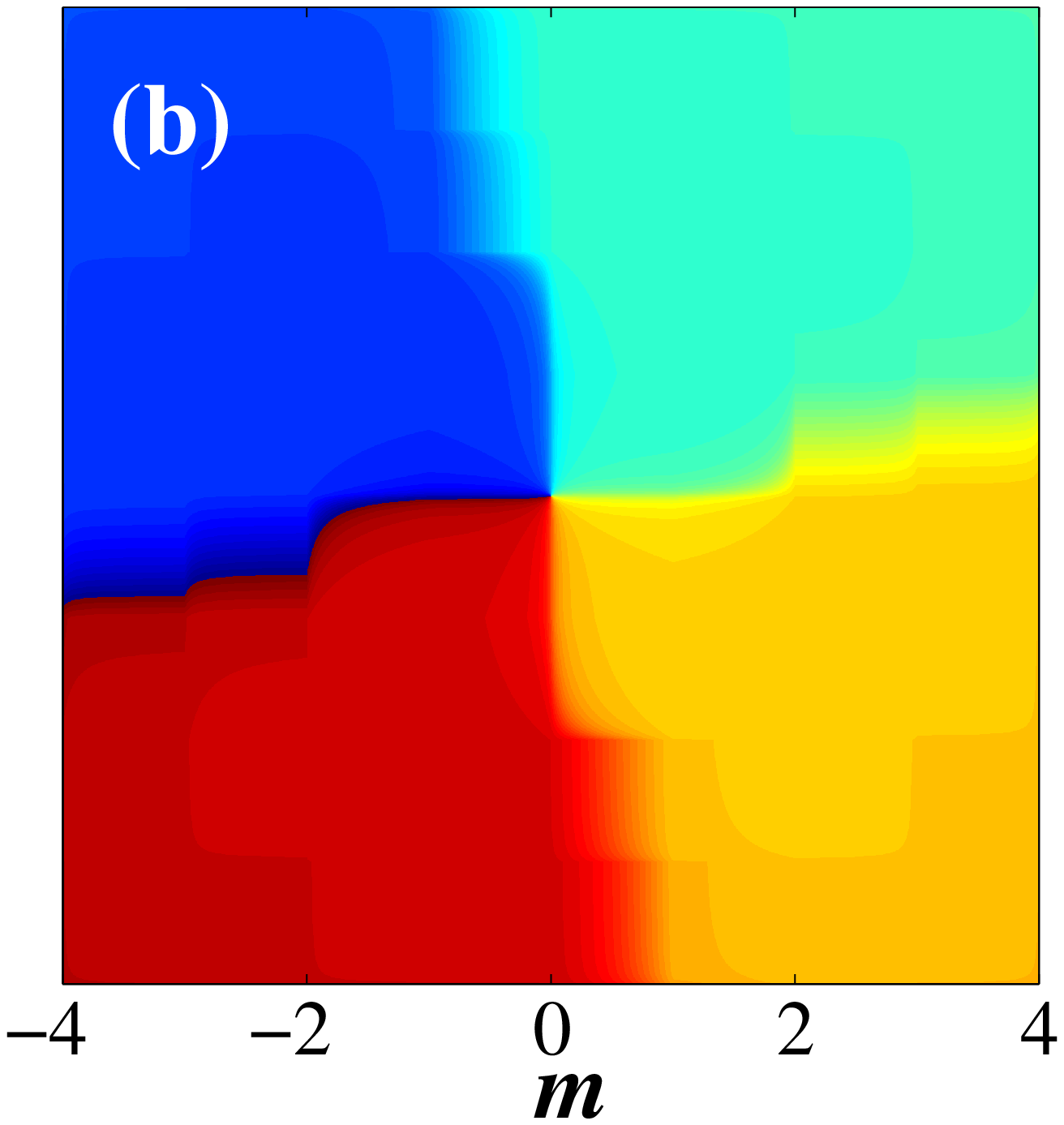,width=0.285\linewidth,clip=}&
\epsfig{file=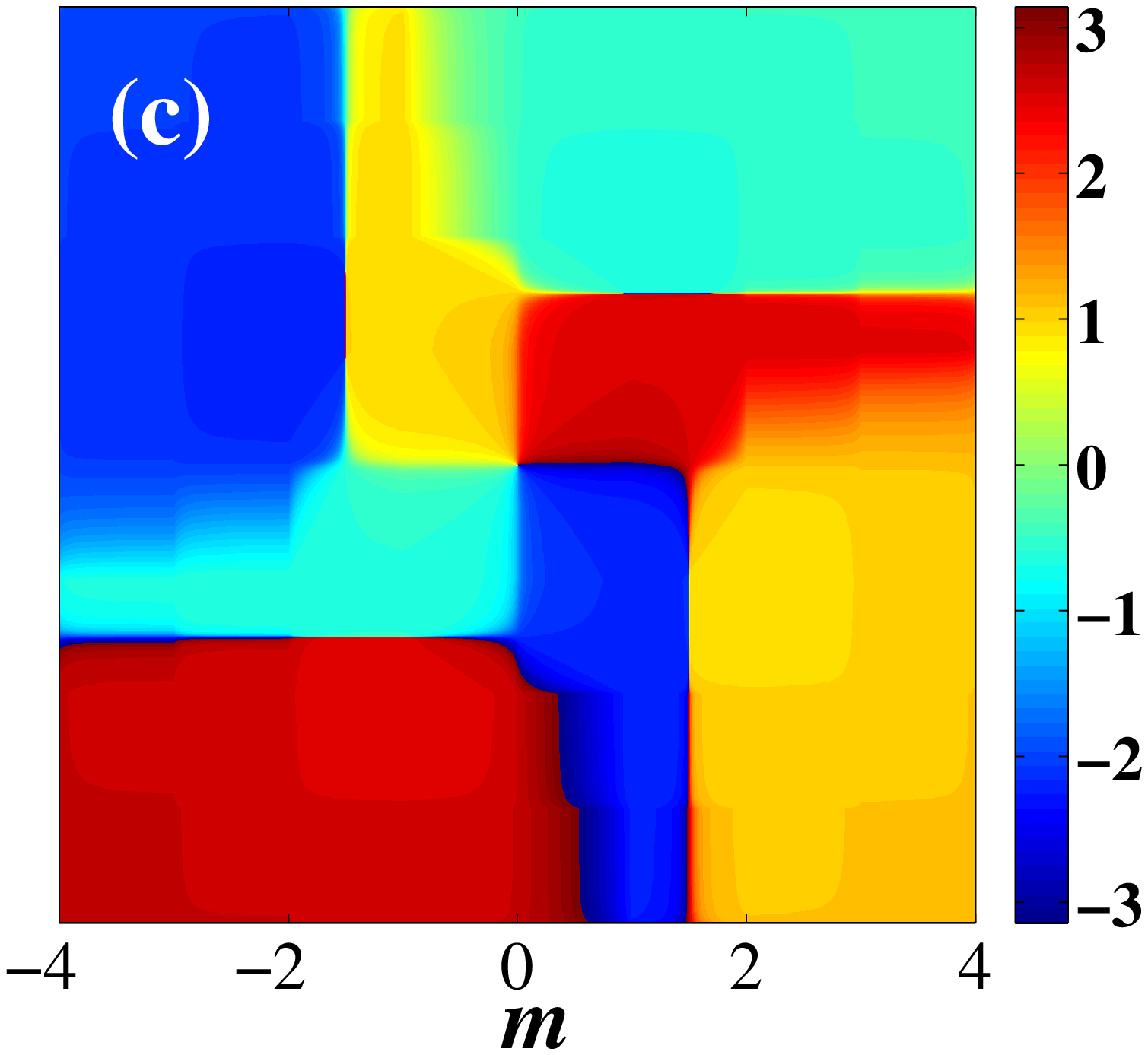,width=0.33\linewidth,clip=}
\end{tabular}
\caption{(Color online) Color map plots for the solutions indicated with dots in
Fig.\ref{fig2}(a). (a) Amplitude profile. (b) and (c) Phase profiles for the
gray and the black points, respectively.}
\label{fig3}
\end{figure}

The amplitude profile for solutions corresponding to the gray and black points
at Fig.\ref{fig2}(a) is shown in Fig.\ref{fig3}(a). This profile is almost
identical for both solutions and it corresponds to a new structure that we
define as ``\textit{swirl-vortex soliton}". However, both solutions have a quite
different phase profile. The unstable solution (belonging to branch $D$)
possesses a full phase profile with charge $S=1$ [see Fig.\ref{fig3}(b)]. A very
interesting thing related with charges happens with the stable swirl-vortex
soliton [see Fig.\ref{fig3}(c)]. For the first square contour [the innermost
discrete square trajectory on the plane $(n,m)$]  we can see that the
vorticity has a $S=1$ value, while for the next contours the vorticity has
decreased to $S=-3$ [$S>0$ ($S<0$) means a clockwise phase structure from $-\pi$
to $\pi$ ($\pi$ to $-\pi$)]. Therefore, there is a stable coexistence of two
different topological charges for the same mode. This new type of structure
would correspond to a ``\textit{two-charges swirl-vortex soliton}'' and - as far
as we know - this would be the first time they are predicted in nonlinear
lattices.

\begin{figure}[h]
\centering
\begin{tabular}{ccc}
\epsfig{file=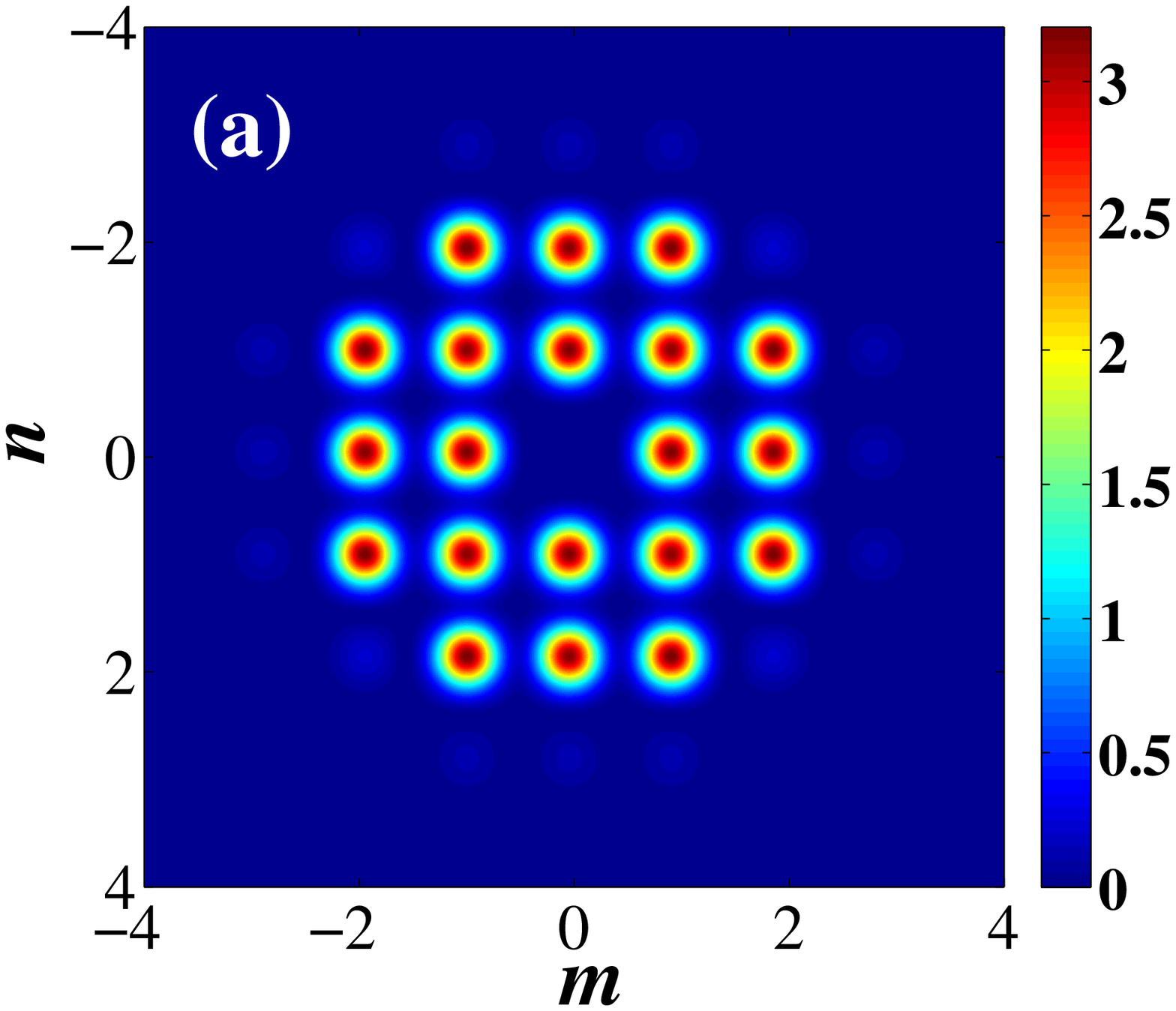,width=0.37\linewidth,clip=}&
\epsfig{file=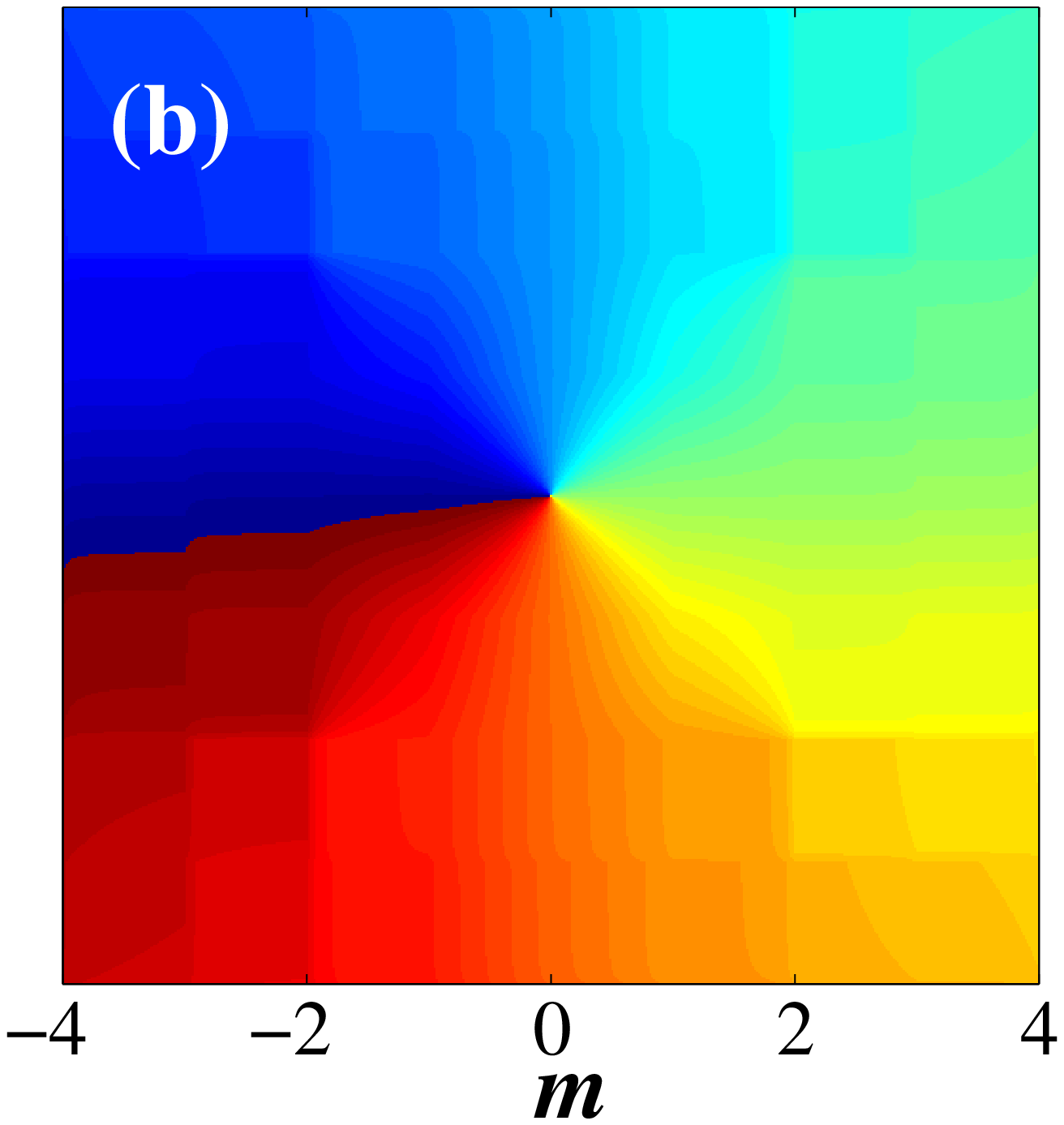,width=0.285\linewidth,clip=}&
\epsfig{file=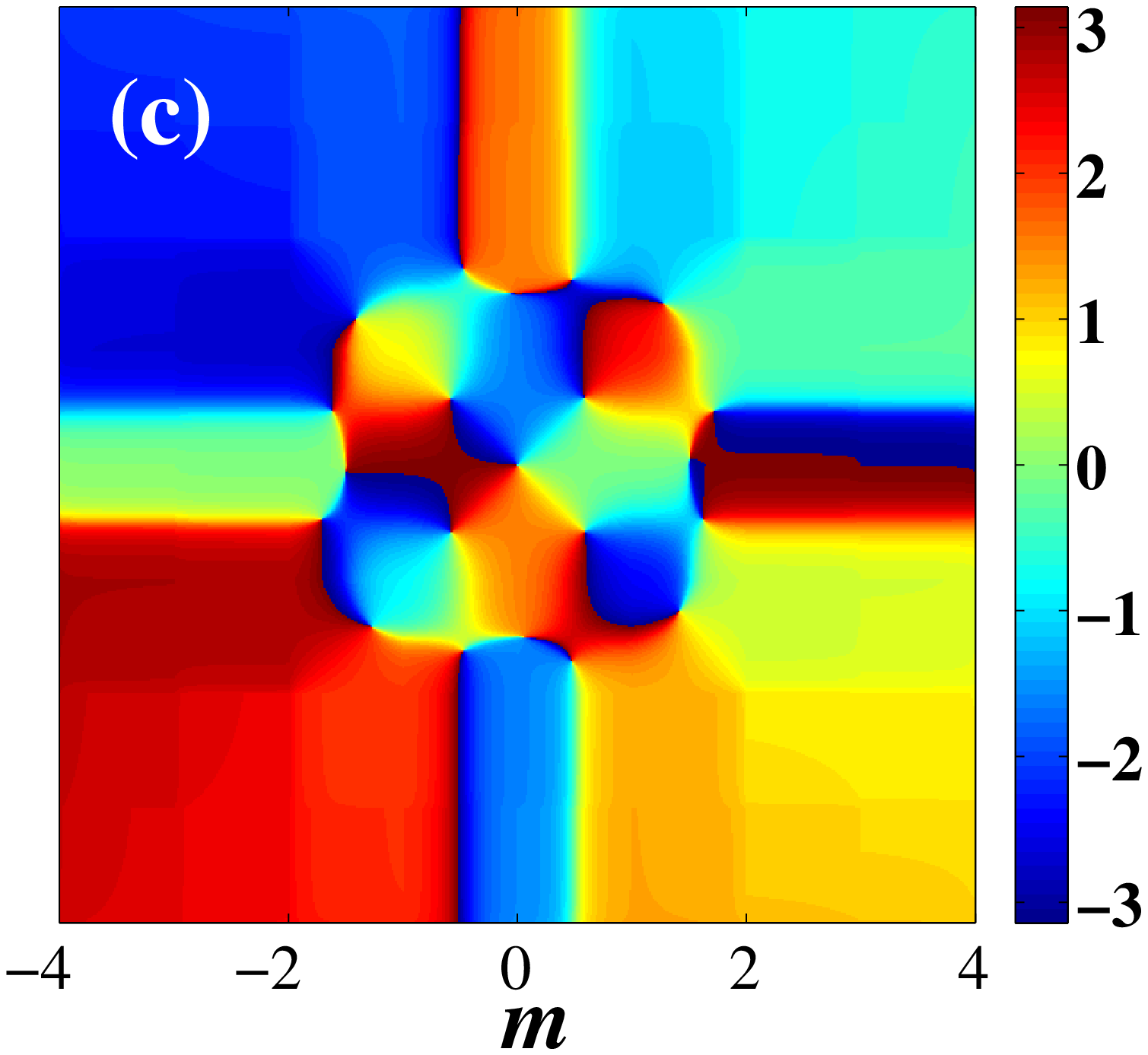,width=0.33\linewidth,clip=}
\end{tabular}
\caption{(Color online) Color map plots for the twenty peaks vortex solution.
(a) Amplitude profile. (b) and (c) Phase profiles for the unstable $S=1$ and the
stable $S=-3$ solutions, respectively. $C=0.8$, $\delta=-0.9$, $\mu=-0.1$,
$\nu=0.1$ and $\varepsilon=1.1$.}
\label{fig4}
\end{figure}

In order to go deeper in the understanding of this stabilization process and its
dependence with the phase structure, we found another example in which the
change in vorticity is also related to the stabilization of the solution. As an
initial {\em ansatz}, we constructed (in the same way as the four-peaks vortex)
a symmetrically-centered twenty peaks $S=1$ configuration. From the
Newton-Raphson scheme we obtain an unstable vortex solution with the amplitude
and phase profiles shown in Figs.\ref{fig4}(a) and (b), respectively. Again, we
use this solution as an initial condition and numerically integrate
Eq.(\ref{dgl2}). As a consequence of the larger number of excited sites, the
power of this solution is higher, namely $Q= 202.43$. Similar to the
swirl-vortex case, the solution evolves and converges to a stable solution with
a very similar amplitude profile, but with a very different phase structure. For
this case, the topological charge of the solution transforms from $S=1$ into
$S=-3$. It is worth to mention that there is a mismatch between the vorticity of
the first and next contours of the lattice [see Fig.\ref{fig4}(c)].

Looking at the colormaps for the stable vortex soliton shown in
Figs.\ref{fig1}(b) and (c) we can realize that amplitude and phase structures
have the same reflection and rotation symmetries. Similar happens with
Figs.\ref{fig3} and Figs.\ref{fig4}. As it was shown in previous works the
stability for one solution with a high number of excited sites requires an
increment of its topological charge~\cite{kevre2005, terh}. From this we
understand why the dynamic evolution modify the vorticity of our solutions. 
So,we may conclude that the instability for complex-structure solutions of charge
$S=1$ is essentially related with the geometric distribution and the number of
excited sites. 

\begin{figure}
\centering
\begin{tabular}{cccc}
\epsfig{file=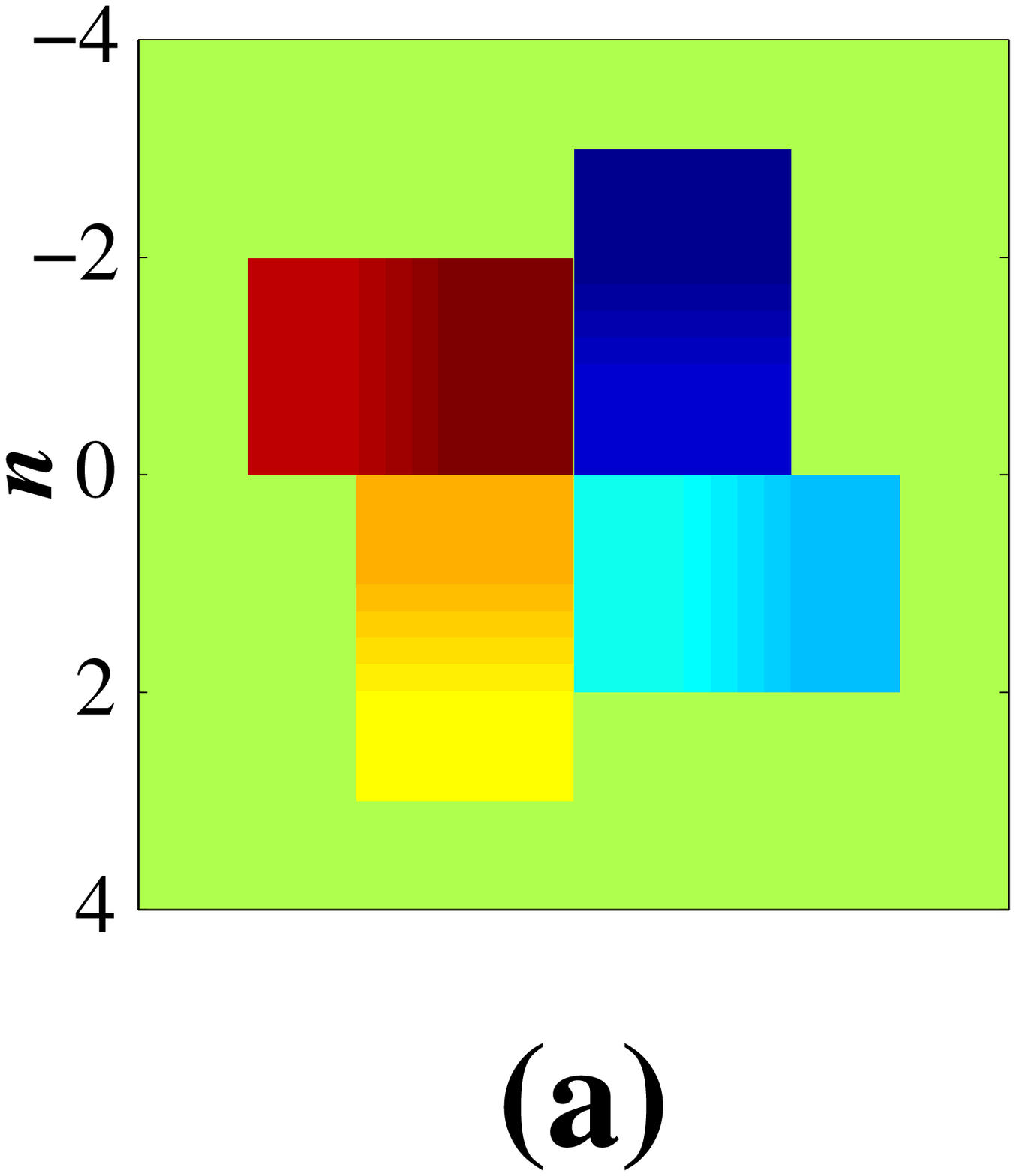,width=0.225\linewidth,clip=}&
\epsfig{file=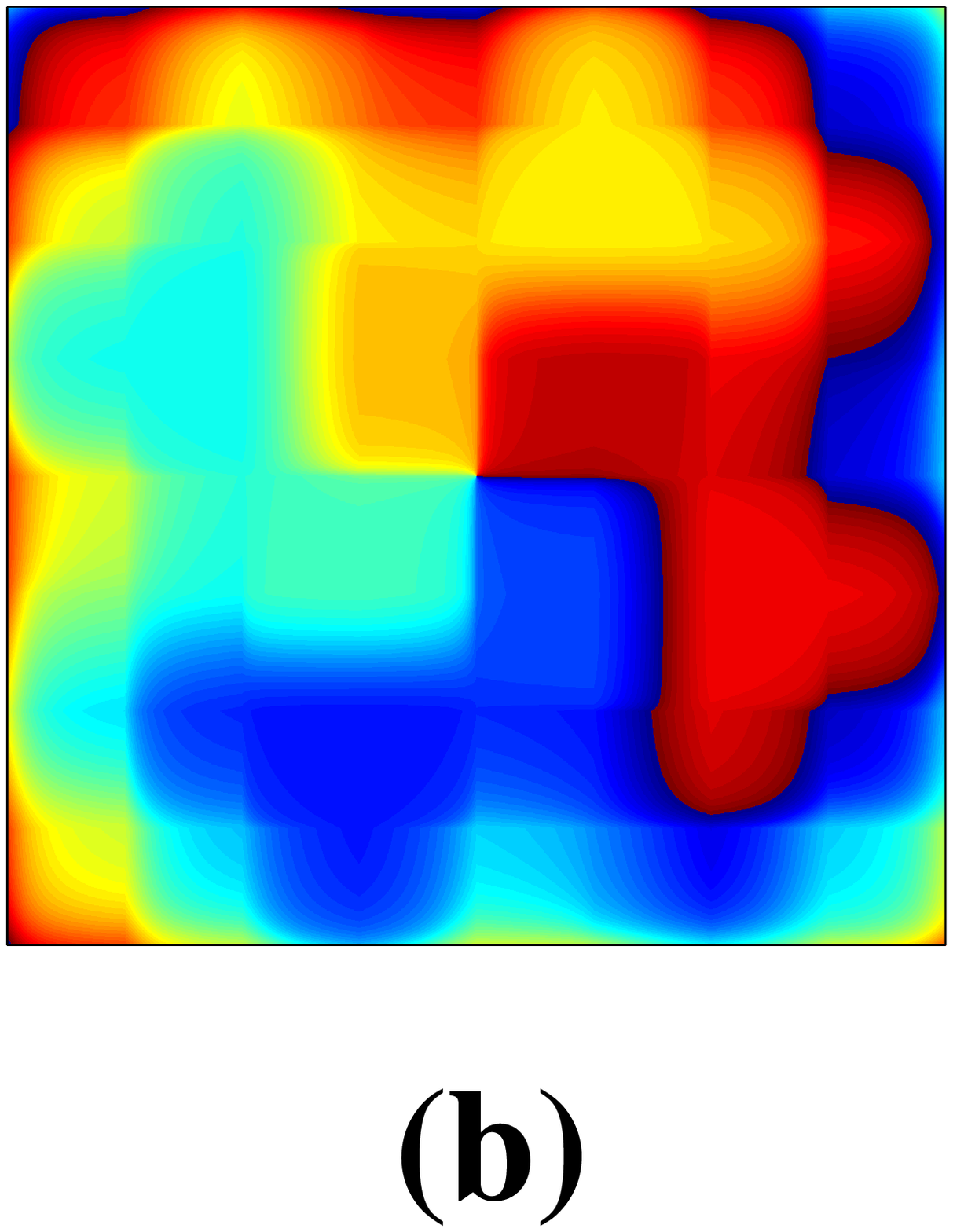,,width=0.198\linewidth,clip=}&
\epsfig{file=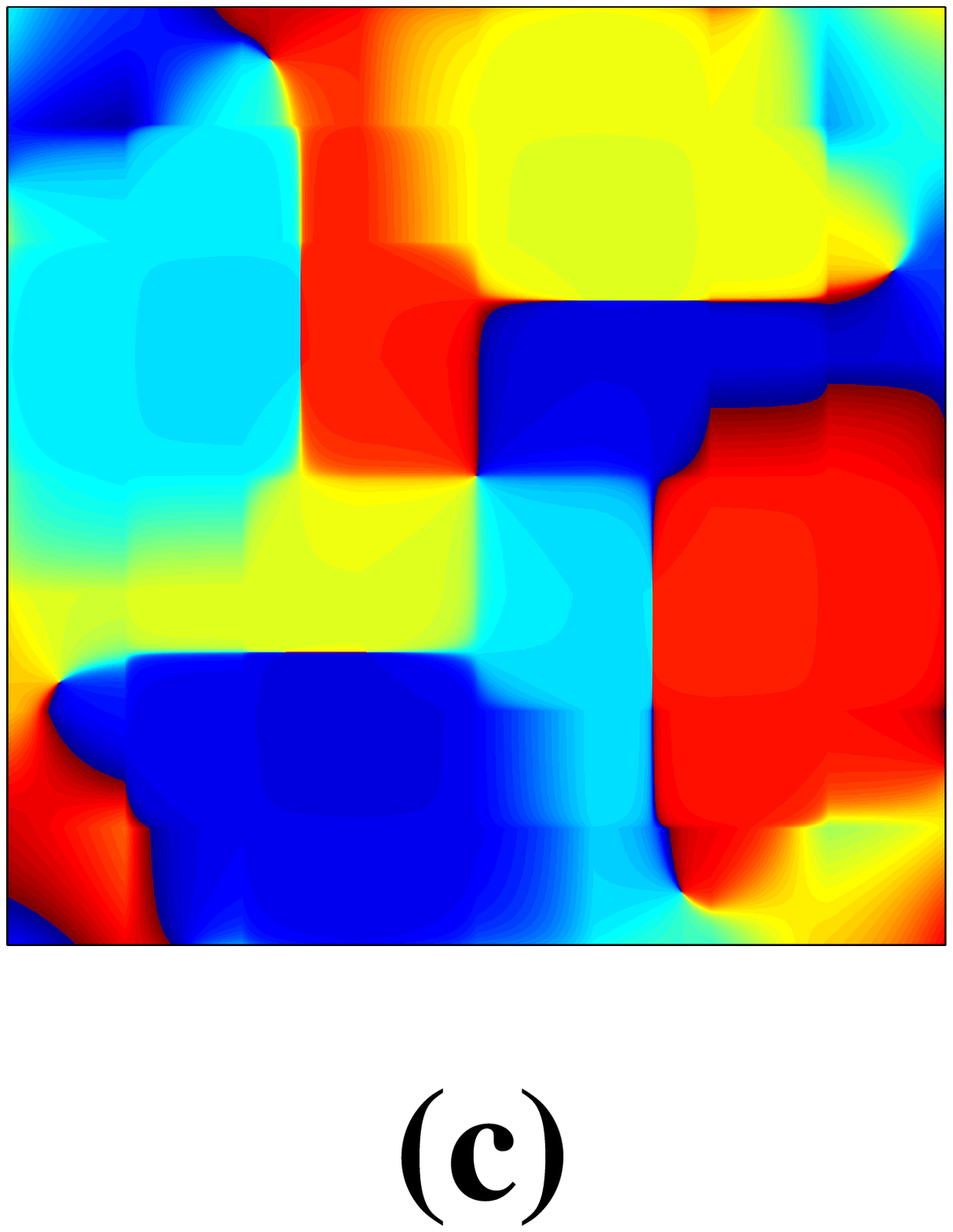,,width=0.198\linewidth,clip=}&
\epsfig{file=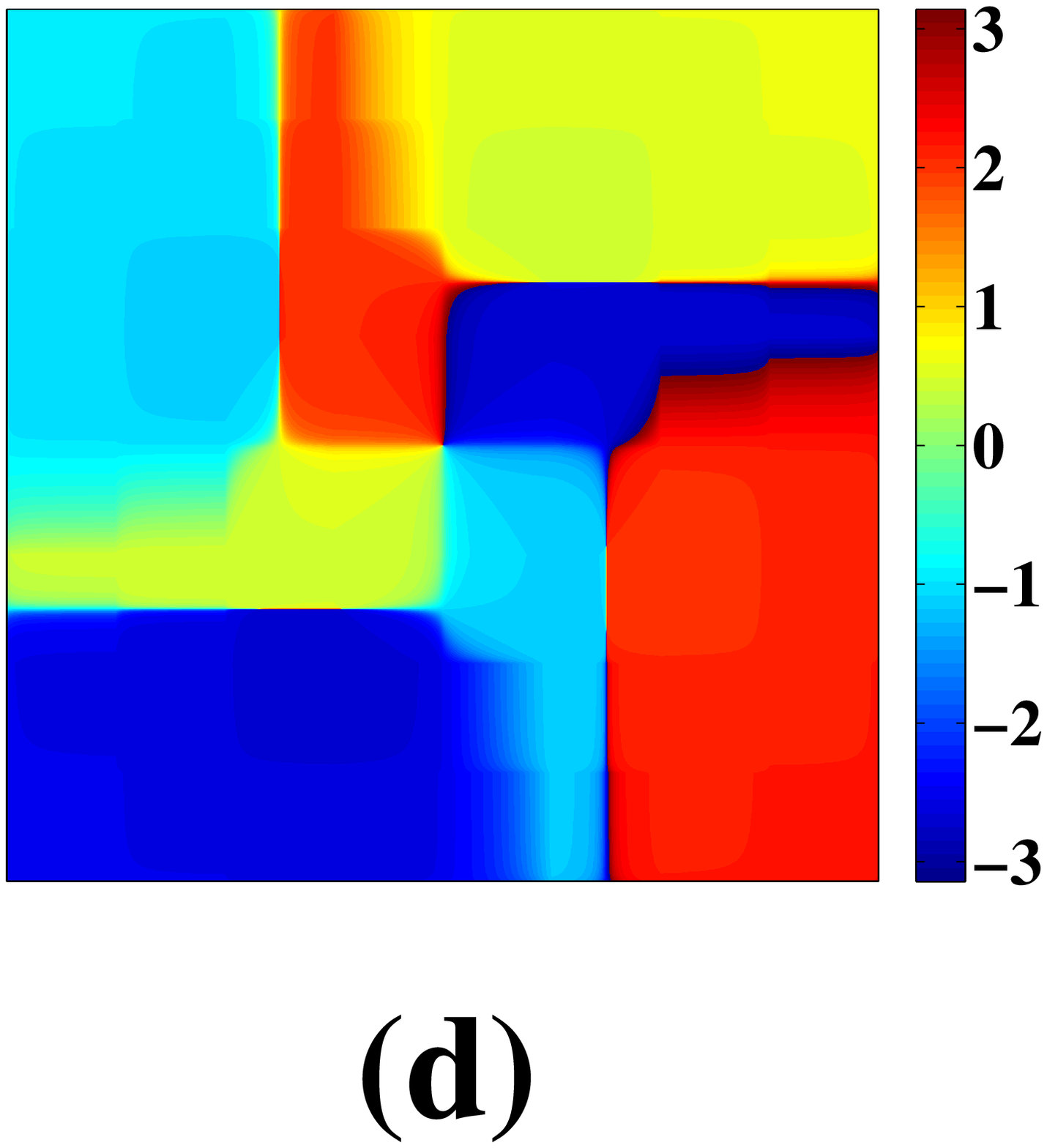,,width=0.235\linewidth,clip=}\\
\epsfig{file=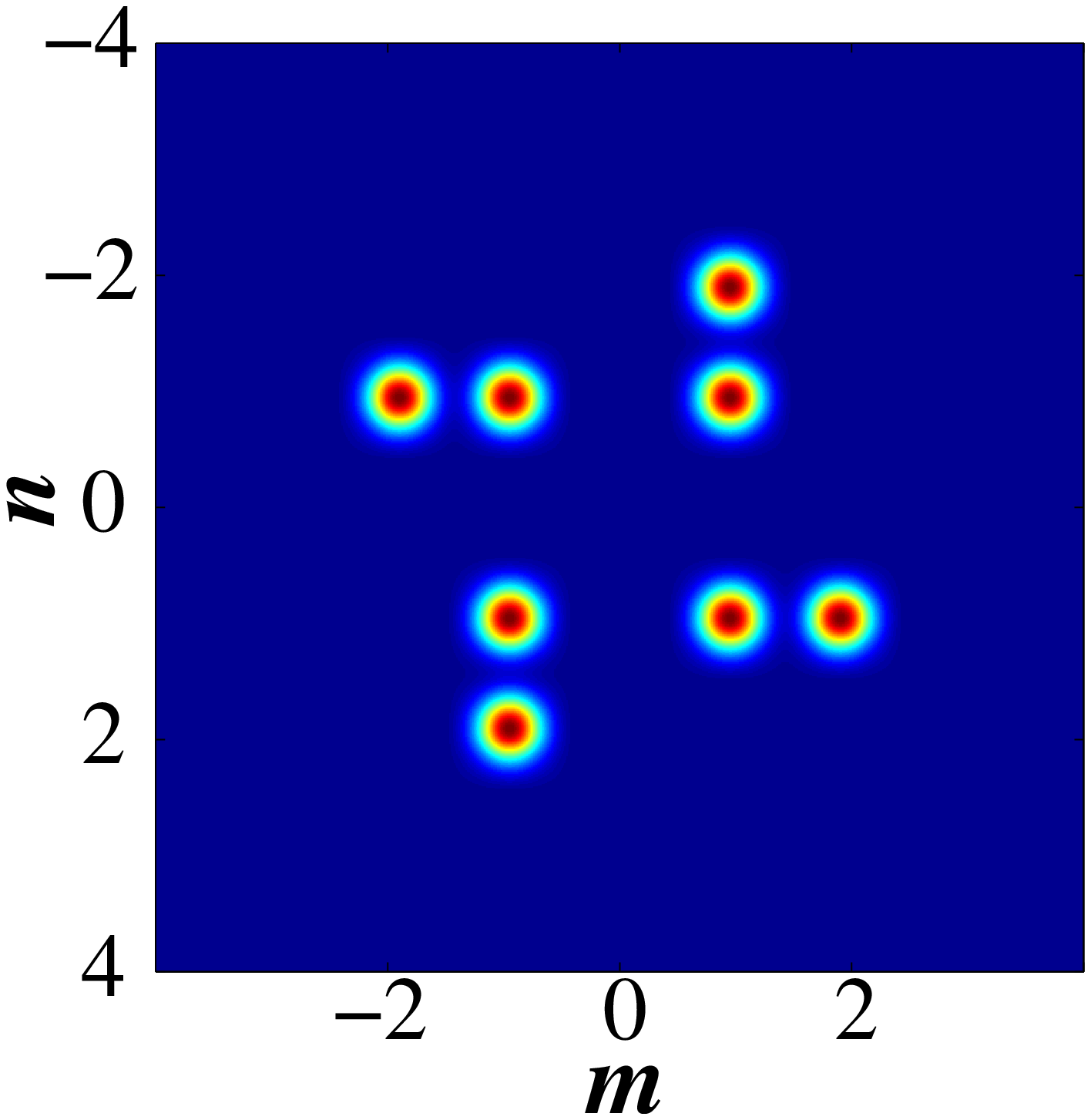,width=0.25\linewidth,clip=}&
\epsfig{file=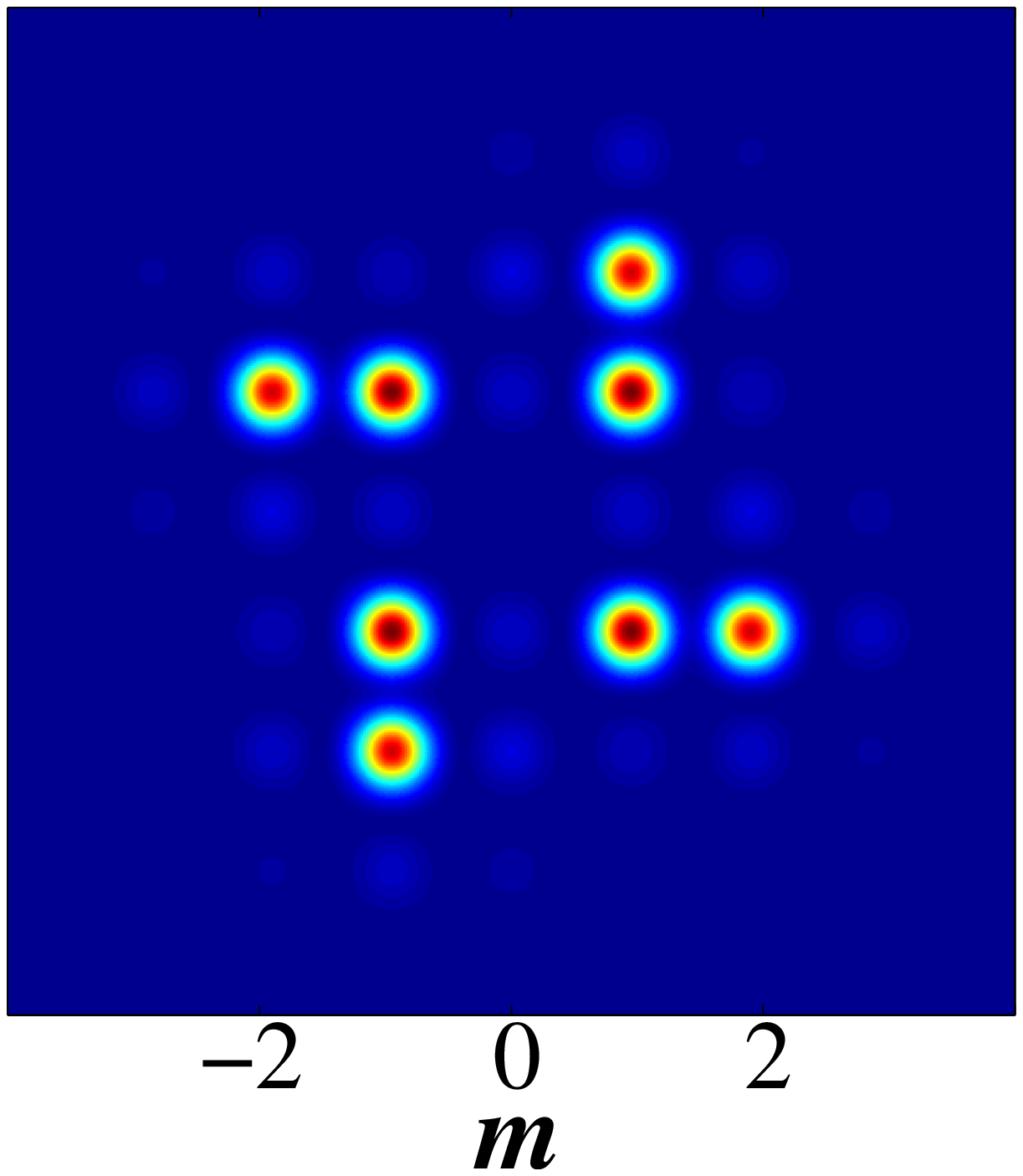,,width=0.223\linewidth,clip=}&
\epsfig{file=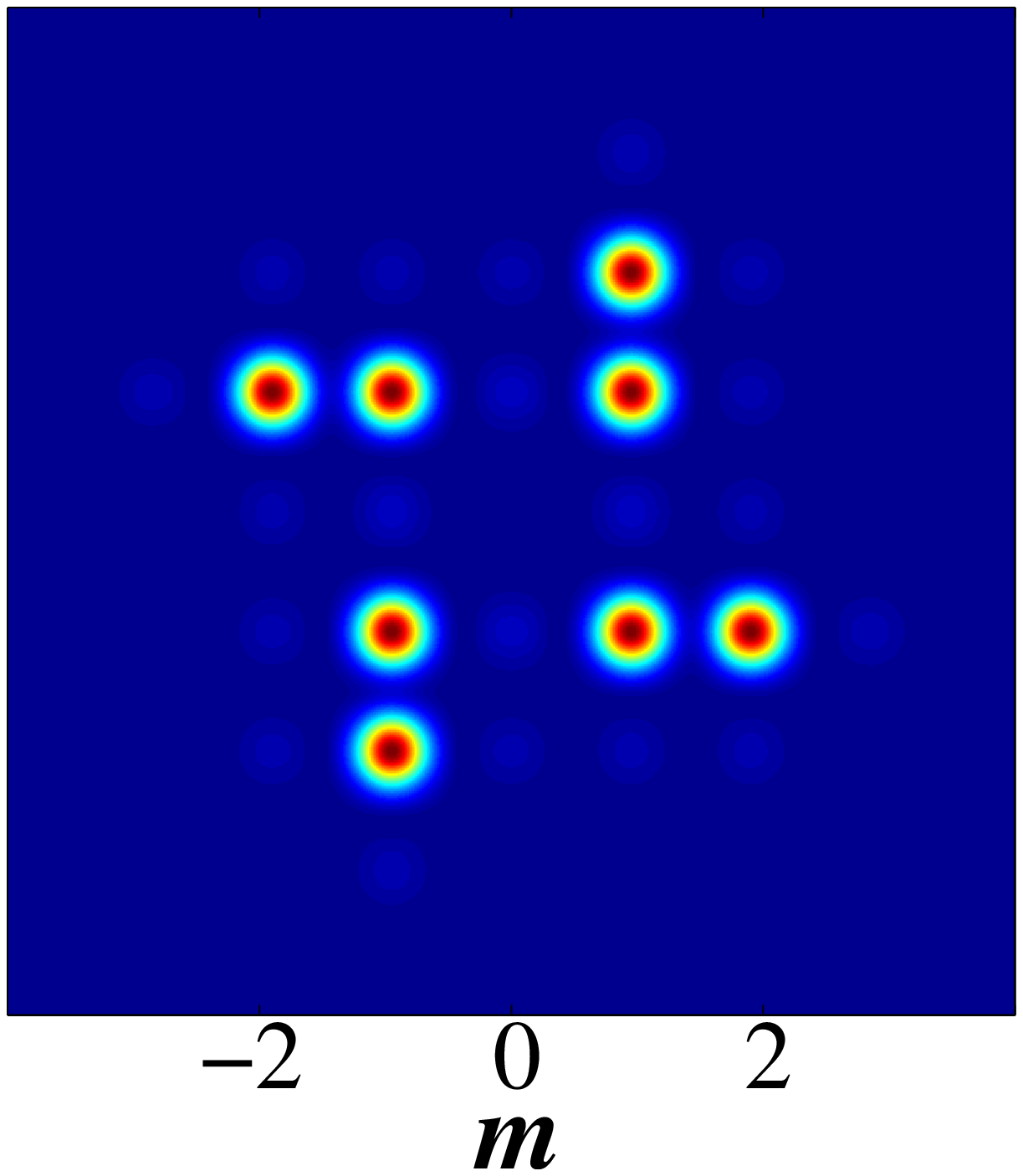,,width=0.223\linewidth,clip=}&
\epsfig{file=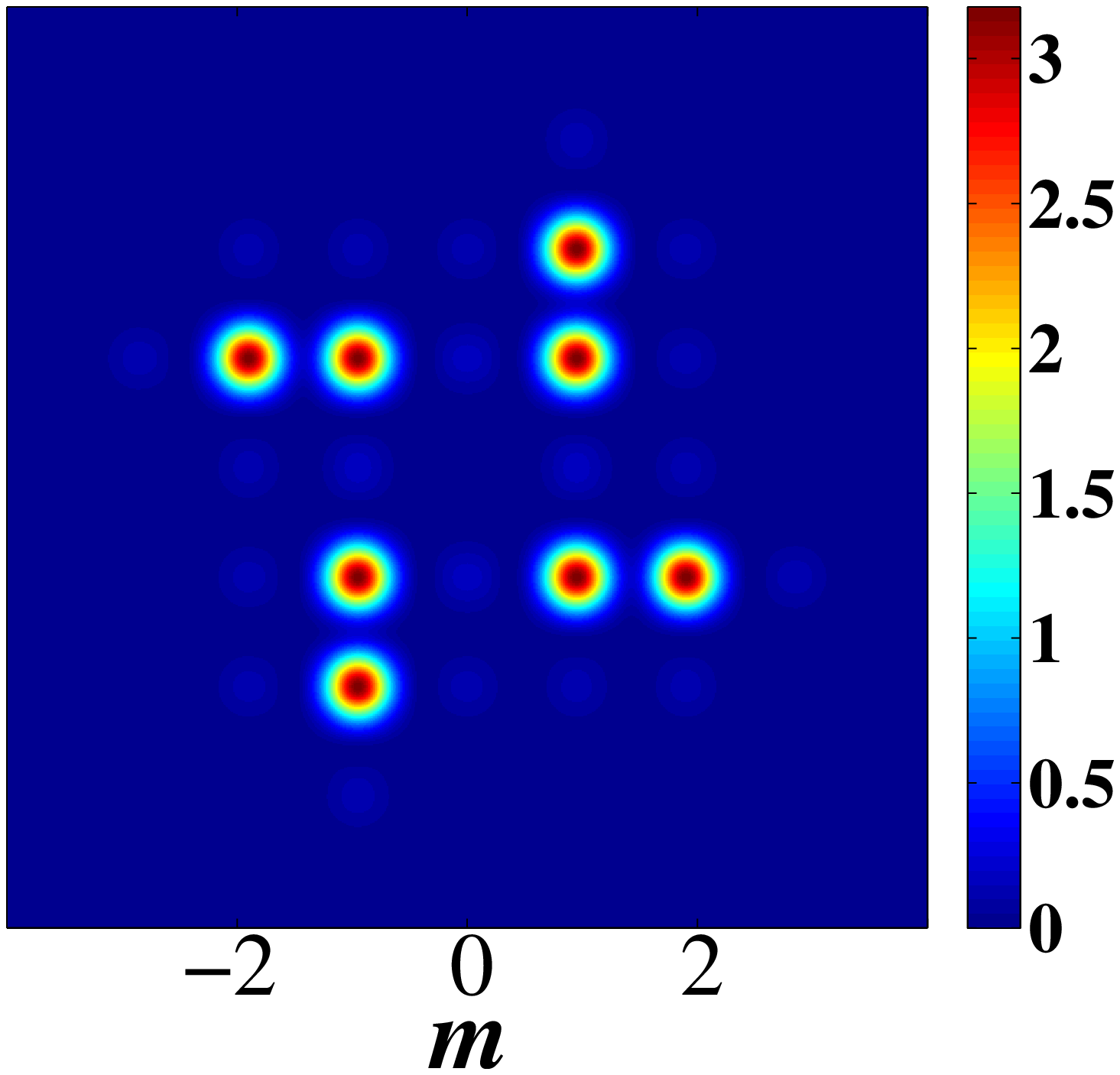,,width=0.26\linewidth,clip=}
\end{tabular}
\epsfig{file=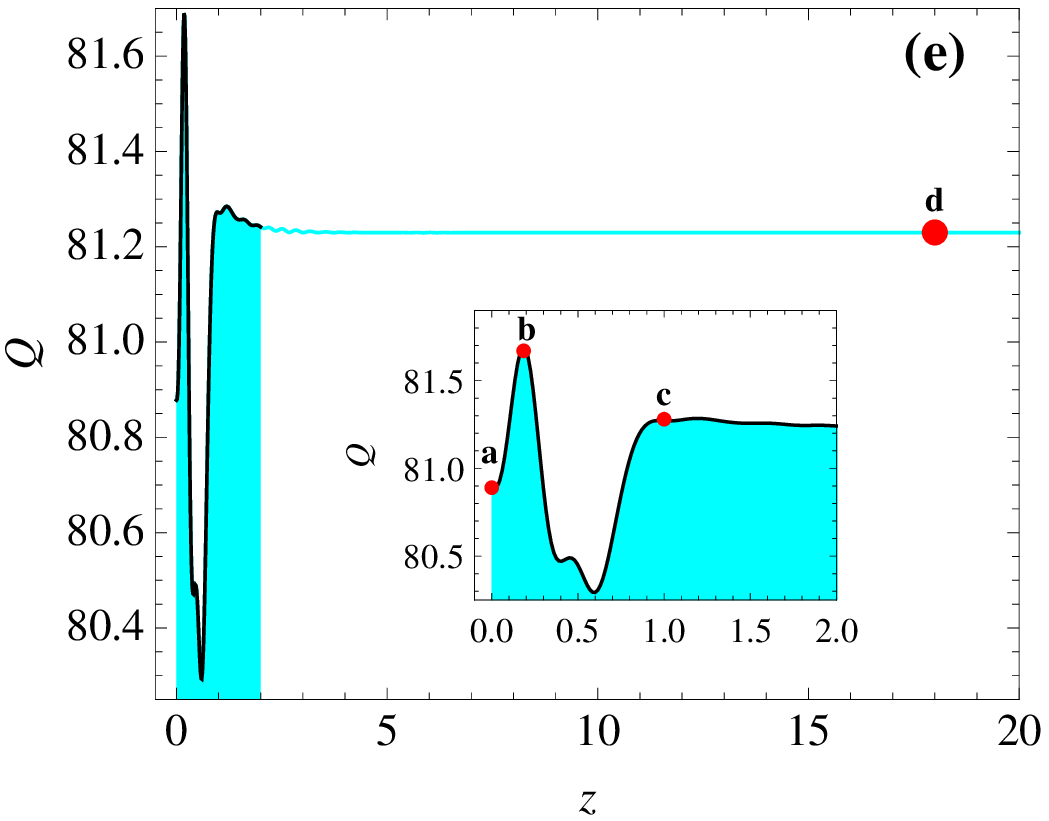,width=0.8\linewidth,clip=}
\caption{(a), (b), (c) and (d) Phase (upper) and amplitude (lower) profiles for different
values of $z$ marked by red points in (e) where the power evolution is displayed.}
\label{fig5} 
\end{figure}

We have also explored the above phenomenology in conservative-cubic systems
(DNLS limit): $\delta=\mu=\nu=\varepsilon=0$. There, we found two branches of
swirl-vortex solitons; one with charge $S=1$ and one with ``two-charges''. The
first one is always unstable while the two-charges solution is only stable for
higher level of power. Here, we did not observe any decaying mechanism
essentially because the system has not gain to increase the power and stabilize
the solution.

Finally, and with the aim of proposing a possible experimental realization, we
numerically integrate model (\ref{dgl2}) by taking - as an initial condition - a
profile with eight peaks spatially distributed in the form of a swirl-vortex,
including its phase structure $S=1$, as Fig.\ref{fig5}(a) shows. With this
configuration we initialize the dynamic evolution observing that the system
rapidly converges to an stable stationary \textit{two-charges swirl-vortex
soliton} [see Figs.\ref{fig5}(a)-(d)]. We can see that the amplitude profiles
slightly change during the propagation: the eight initial peaks remain almost
unaltered. On the other hand, from Figs.\ref{fig5}(d) we see that for the first
square contour the vorticity is preserved being $S=1$, while for the next
contours the charge has transformed into $S=-3$. Figs.\ref{fig5}(e) shows the
evolution of power with some initially small oscillations and, lately, a
tendency to the stabilization of the profile. This example shows the robustness
of our prediction and its chances to be observed in real dissipative systems
because the initial condition could, in principle, be easily implemented in
current experimental setups. Another very interesting point is that the system
naturally evolves to a ``2-charges'' structure. Our initial condition has an
unstable phase structure which guarantees the decaying to another type of mode,
but not necessarily to the one we are interested in; it could perfectly just be
destroyed by the internal dynamics~\cite{contiperio}. However, the system favors
the excitation of a swirl-vortex solution which propagates stably for long
propagation distances.

In conclusion, our results reveal the existence of discrete vortex solitons in
dissipative 2D-lattices. We have found stable and unstable vortices by
performing different continuation methods. In particular we concentrated the
study on a new type of stable structure, the so-called \textit{two-charges
swirl-vortex soliton}. We were able to dynamically excite it by using a simple
initial configuration and therefore, believe in the feasibility of experimental
observation of this novel type of dissipative structures.

C.M.C. and J.M.S.C. acknowledge support from the Ministerio de Ciencia e
Innovaci\'on under contracts FIS2006-03376 and FIS2009-09895. R.A.V and M.I.M
acknowledge support from FONDECYT, Grants 1080374 and 1070897, and from 
Programa de Financiamiento Basal de CONICYT (FB0824/2008).


\begin{thebibliography}{99}

\bibitem{PT}D.K. Campbell, S. Flach, and Yu.S. Kivshar, Phys. Today {\bf 57} (1), 43 (2004).

\bibitem{rep1} F. Lederer, G.I. Stegeman, D.N. Christodoulides, G. Assanto, M. Segev, and Y. Silberberg, Phys. Reps. \textbf{463}, 1 (2008).

\bibitem{rep2}S. Flach and A. Gorbach, Phys. Reps. {\bf 467}, 1 (2008).

\bibitem{chrinat}D.N. Christodoulides, F.Lederer, and Y.Silberberg, Nature (London) {\bf 424}, 817 (2003).

\bibitem{heis}H.S. Eisenberg, Y.Silberberg, R.Morandotti, A.R. Boyd, and J.S. Aitchison, \prl {\bf 81}, 3383 (1998).

\bibitem{fleis}J.W. Fleischer, M. Segev, N.K.Efrimidis, and D. N. Christodoulides, Nature (London) {\bf 422}, 147 (2003).

\bibitem{vortex1}D.N. Neshev et al., \prl {\bf 92}, 123903 (2004); J.W. Fleischer et al., \prl{\bf 92}, 123904 (2004).

\bibitem{opex09}J.M. Soto-Crespo, N. Akhmediev, C. Mej\'ia-Cort\'es, and N. Devine, Opt. Exp. {\bf 17}, 4236 (2009).

\bibitem{contiperio}H. Leblond, B.A. Malomed, and D. Mihalache, \pra {\bf 80}, 033835 (2009).

\bibitem{soto}J.M. Soto-Crespo, N. Akhmediev, and G. Town, \oc {\bf 199}, 283 (2001).

\bibitem{akhm0508}
N. Akhmediev and A. Ankiewicz,
\newblock {\em Dissipative Solitons: From optics to biology and medicine},
\newblock Springer, New York, 2005;

\bibitem{kevre2005}D. Pelinovsky, P. Kevrekidis, and D. Frantzeskakis, Physica D {\bf 212}, 20 (2005).

\bibitem{efre07}N.K. Efremidis, D.N. Christodoulides, and K. Hizanidis, \pra {\bf 76}, 043839 (2007).

\bibitem{kivshar2005}T.J. Alexander, A.A. Sukhorukov, and Y.S. Kivshar, \prl {\bf 93}, 063901 (2004).

\bibitem{stabi}R.A. Vicencio and M. Johansson, \pre {\bf 73}, 046602 (2006).

\bibitem{terh}Bernd Terhalle, Tobias Richter, Kody J. H. Law et al., \pra {\bf 79}, 043821 (2009).

\end{thebibliography}
\end{document}